\documentstyle[aps,multicol]{revtex}

\newcommand{\be}{\begin{equation}} 
\newcommand{\ee}{\end{equation}}
\newcommand{\bea}{\begin{eqnarray}} 
\newcommand{\eea}{\end{eqnarray}}

\def\l{\lambda}

\def\IZ{\relax\ifmmode\mathchoice
{\hbox{\cmss Z\kern-.4em Z}}{\hbox{\cmss Z\kern-.4em Z}}
{\lower.9pt\hbox{\cmsss Z\kern-.4em Z}}
tn
{\lower1.2pt\hbox{\cmsss Z\kern-.4em Z}}\else{\cmss Z\kern-.4em Z}\fi}

\begin{document}

\title{Renormalization group irreversible functions in more than
two dimensions} 

\author{Jos\'e Gaite%
\footnote{Also at {\it Instituto de Matem{\'a}ticas y F{\'\i}sica
Fundamental, CSIC, Serrano 123, 28006 Madrid, Spain}}
}

\address{Centro de
Astrobiolog{\'\i}a, Instituto Nacional de T\'ecnica
Aeroespacial,\\
Ctra.\ de Torrej\'on a Ajalvir,
28850 Torrej\'on de Ardoz, Madrid, Spain.
}

\date{August 24, 2000}
\maketitle

\begin{abstract} 
There are two general irreversibility theorems for the renormalization
group in more than two dimensions: the first one is of entropic nature,
while the second one, by Forte and Latorre, relies on the properties of
the stress-tensor trace, and has been recently questioned by Osborn
and Shore. We start by establishing under what assumptions this second
theorem can still be valid. Then it is compared with
the entropic theorem and shown to be essentially equivalent. However,
since the irreversible function of the (corrected) Forte-Latorre
theorem is non universal (whereas the relative entropy of the other
theorem is universal), it needs the additional step of
renormalization. On the other hand, the irreversibility 
theorem is only guaranteed to be unambiguous if 
the integral of the stress-tensor trace correlator is finite, which happens 
for free theories only in dimension smaller than four. 
\end{abstract}


\begin{multicols}{2}
The search for a function representing the irreversible nature of the
coarse-graining transformations of Wilson's renormalization group (RG)
has a long history. After the success of Zamolodchikov's $c$-function
in two dimensions ($2D$), it was shown that a straight-forward
generalization to higher dimension was not possible but, at the same
time, it was observed that a related function, the integral of the
stress-tensor trace on a constant curvature space, could play a
similar role \cite{Cardy}.  In an interesting article \cite{FoLa}
Forte and Latorre formulated an irreversibility theorem in terms of
this quantity.  However, an exhaustive analysis of this theorem
carried out by Osborn and Shore \cite{OsbS} shows that there were
missing terms in that theorem that actually spoil the irreversible
character of that function.

In a separate development, we have introduced in Field Theory
the relative entropy, a quantity 
borrowed from probability theory 
which turns out to be the Legendre
transform of $W(\l)-W(0)$ with respect to $\l$ \cite{I-OC}:
\bea 
S_{\rm rel}(\l) &=& W(\l) - W(0) - \l{d W\over d\l}\nonumber\\
&=& W - W_0 - \l\,\langle f_\l \rangle, \label{Legtrans} 
\eea 
where $f_\l$ is a {\em composite field} integrated over the whole
space, $f_\l = \int d^D\!x \,\Phi(x)$.  As a straightforward
consequence of its definition, the relative entropy satisfies a
monotonicity theorem,
\bea 
{dS_{\rm rel}\over d\l} &=& {dW\over d\l} -
{d\over d\l}\left(\l{dW\over d\l}\right) =- \l {d^2W\over d\l^2}\nonumber\\
&=&- \l{d\over d\l}\langle f_\l\rangle =
\l\,\langle (f_\l-\langle f_\l\rangle)^2\rangle \geq 0,
\label{pos}
\eea
which can be interpreted as showing the irreversibility of the RG
\cite{I-OC,I}. As we remarked in previous papers, the relative entropy
is not the only monotonic quantity with the RG. For example, from the
same equation that shows its monotonicity (\ref{pos}), one can
realize that the function $\langle f_\l\rangle = {dW/d\l}$ is
monotonic as well. 

Let us consider the integral $\int d^D\!x\,\langle\Theta(x) \rangle$,
where $\Theta$ is the stress tensor trace.  In a homogeneous space the
expectation value $\langle \Theta \rangle$ is independent of the
position and only depends on the coupling constants; hence, the
integration is trivial, its only effect being to produce an overall
factor.  We further consider a field theory with {\em simple scaling
behavior}, namely, with only one coupling constant such that $\l
\propto m^{y}$, where $m$ is the physical mass of the fundamental
particle or some other mass scale.  This behavior is very common in
critical phenomena.  Since $\Theta$ gives the response to a
change of the scale $m$,
\be
\langle \Theta \rangle \equiv m{d W\over dm} = y\,\l{d W\over d\l} = 
y\,\l\,\langle \Phi \rangle,
\ee
where now $W$ is a specific quantity (per unit volume).  In other
words, the expectation value of $\Theta$ is proportional to the
monotonic function $\langle f_\l\rangle$. By substituting for it in
the monotonicity equation (\ref{pos}), we can write this equation in
the maybe more suggestive form
\be
- m{d\over dm}(m^{-y}\,\langle\Theta \rangle) =
m^{-y}\int\! d^D\!x\,\langle\Theta(x)\Theta(0)\rangle_c,
\label{pos2}
\ee
where the subscript $c$ means that one is to take the {\em connected}
correlation function.  The integral of this correlation function may
be divergent.  If $m \neq 0$ it converges for $x \rightarrow
\infty$. On the other hand, the behavior of the two-point function for
$x \rightarrow 0$ is the same as in the massless ($\l=0$) theory, thus
given by the dimension of $\Phi$, $d_\Phi$. Therefore, the integral is
UV convergent if $2\,d_\Phi < D$, that is, if $y = D - d_\Phi > D/2$.
One can also derive an equation for $\langle \Theta \rangle$:
\be
- m{d\over dm}\langle\Theta \rangle =
\int\! d^D\!x\,\langle \Theta(x)\Theta(0)\rangle_c 
- y\,\langle\Theta \rangle.
\label{Osb}
\ee

In Euclidean space the form of the quantities defined above is given
by scaling (e.g., $S_{\rm rel} \propto m^D$) and has little physical
content. In a curved homogeneous space one can form the dimensionless
variable $u=R\,m$, where $R$ is the curvature radius, and
dimensionless quantities are non-trivial functions of it. In
particular, we have the dimensionless function of
Refs.~\cite{Cardy,FoLa,OsbS}, $c(u) = R^{D}\,\langle\Theta \rangle$.
Introducing a constant curvature space has an additional utility:
Eq.~(\ref{Osb}) can also be obtained starting from the scale Ward
identities satisfied by the energy momentum tensor as $R$ varies
\cite{FoLa,OsbS}.  Let us remark that the derivation in
Ref.~\cite{FoLa} yields a slightly different equation. It has been
polished in Ref.~\cite{OsbS} similar but more general than
Eq.~(\ref{Osb}):
\bea
- R{d\over dR}\left(R^{D}\langle\Theta \rangle\right) &=&
R^{D}\int\! d^D\!x\,\langle \Theta(x)\Theta(0)\rangle_c \nonumber\\
&&{}-R^{D}\beta^i\,(\partial_i{\cal A} + 
\partial_i\beta^j\langle\Phi_j\rangle).
\label{OsbR}
\eea
This equation takes into account the possibility of several couplings and the
existence of the {\em trace anomaly} ${\cal A}$, such that $T_a^a =
\Theta + {\cal A}$, where $\Theta = \beta^i\Phi_i$. We can convert
Eq.~(\ref{OsbR}) into Eq.~(\ref{Osb}) by (i)
assuming simple scaling behavior, that
is, with only one coupling such that $\beta = y\,\l$, the anomaly
${\cal A}$ being independent of it, and by (ii) replacing the derivative with
respect to $R$ with a derivative with respect to $m$.  

Therefore, even though in the general case no monotonicity theorem
seems to follow from Eq.~(\ref{OsbR}) \cite{OsbS}, in our case it
does, namely, the one expressed by Eq.~(\ref{pos2}). However, the
monotonic quantity (with respect to $m$ or $R$, indistinctly) is not
just $c(u) = R^{D}\langle\Theta \rangle$, as proposed in
Refs.~\cite{Cardy,FoLa,OsbS}, but rather $\tilde{c}(u) =
u^{-y}\,c(u)$.  They only coincide if $y=0$, that is, when the
coupling constant is dimensionless. 
Generally, the functions $c$ or $\tilde{c}$, involving the composite field
$\Phi$, contain ({\em normal order}) UV divergences.
We can introduce a UV regulator but, given that it can only be removed 
by introducing another scale ({\em renormalization point}), those 
functions are not universal.

To define a finite monotonic function from the stress-tensor trace, 
one has, therefore, 
to perform a subtraction. Let us define the function 
\be
f(u)=-V_{D-1}\,m^{-y}\,R^{D-y}\,\langle\Theta\rangle,
\ee
where $V_{D-1} =
2\pi^{D/2}/\Gamma(D/2)$ is the volume of the unit $(D-1)$-dimensional
sphere.  This function is essentially $\tilde{c}(u)$, except for a
conventional sign (to make it increasing rather than decreasing) 
and a normalization factor. It is
UV divergent but, assuming the convergence of the integral 
in Eq.~(\ref{pos2}), 
one subtraction suffices to render it finite. The
point is that when integrating $df/dm$ according to
Eq.~(\ref{pos2}), one has an integration constant, which can be infinite.
Therefore, we can define a renormalized value as \cite{I2}
\be
f_{\rm ren}(mR) := \lim_{\Lambda\rightarrow\infty}
[f_{\Lambda}(mR) - f_{\Lambda}(m_0 R)],
\ee
${\Lambda}$ and $m_0$ being the UV cutoff and the subtraction point,
respectively. In particular, one can set $m_0=0$. Alternatively, one
can use {\em minimal subtraction}, by which one only subtracts the
divergent part of $f$, which is independent of $m$ \cite{MS}. 
In any renormalization scheme we use the freedom afforded by 
the integration constant of Eq.~(\ref{pos2}), for example,
to make $f(0) = 0$, which is equivalent to taking $m_0 = 0$.  

In contrast, the relative entropy
is universal (under the assumption of convergence of the integral in
Eq.~(\ref{pos2})) because the UV divergences of $W$ cancel in the
definition of $S_{\rm rel}$, Eq.~(\ref{Legtrans}). 
One can define a dimensionless growing entropy $S$, 
proportional to $S_{\rm rel}$. In terms of the function $f$,
\be
S(u) = y\int_0^u dv\, v^{y-1}f(v) - u^{y} f(u).
\label{S}
\ee
The renormalization constant of $f$ cancels in this formula. 

To illustrate the general theory, we will study a free massive
scalar field $\phi$, with coupling constant $m^2$, in $D$-dimensional
hyperbolic space $H^D$, for $D=2,3,4$.  Naturally, a free massive
scalar field theory is the simplest example of simple scaling one can
take.  The field expectation value $\langle\phi^2 \rangle$ is then the
Gaussian model energy $U(m^2)$ ~\cite{I,I2}, while $S_{\rm rel}$ is a real
thermodynamic entropy.  Some expressions for the quantities in $D=2$
have been calculated in Ref.~\cite{I2}, in terms of the variable $r=
(Rm)^2$. More extensive calculations of $\langle\phi^2 \rangle$ are
given by Osborn and Shore \cite{MS}. For $H^2$,
\be 
f_{\rm ren}(r) = \psi(\sqrt{r+1/4}+1/2)+\gamma,
\ee
where $\psi$ is the digamma function and $\gamma$ is the Euler constant. 
$f$ increases with $r$, on account of the properties of $\psi$.

For $D=3$ we could use as well the results of Osborn and Shore
\cite{xi} but it is easier to use instead the heat-kernel method \cite{I2},
since the $D=3$ heat kernel is extremely simple \cite{Campo-rep}:
\be 
K(0;t)={e^{-t}\over (4\pi t)^{3/2}}.
\ee
Hence, 
\bea
{f_{\rm ren}(r)} &=& -4\pi\int\limits_0^\infty {dt\over(4\pi t)^{3/2}}
\left[e^{-(r+1)\,t}-e^{-t}\right]\nonumber\\
&=& \sqrt{r+1} - 1.
\eea
This function is obviously increasing.

In $D=4$, the case used as example in Ref.~\cite{FoLa}, $f$ has an
expression similar to the one for $D=2$ \cite{Campo,MS}.  However, it
is a particularly interesting case because the integral 
in Eq.~(\ref{pos2}) is now
divergent, so $f'_{\Lambda}(r)$ must be subtracted too. 
Consequently, {\em two
subtractions} on $f$ are needed now, that is,
$$f_{\rm ren}(r) = \lim_{\Lambda\rightarrow\infty}[f_{\Lambda}(r) -
f_{\Lambda}(r_0) - (r-r_0)f'_{\Lambda}(r_0)].$$
Subtraction at $m_0 = 0$ yields
\bea
f_{\rm ren}(r) &=& -{1\over 4}
\left[(r + 2)\,\psi(\sqrt{r+9/4}+1/2) \right.\nonumber\\&&{}- 2(1 - \gamma) - 
    \left.(3 - 9\gamma + \pi^2)\,{r\over 9}\right],
\eea
which {\em decreases} for $r > 0$.  The reason is the following. One
can compute $f'_\Lambda(r)$ and it is indeed positive for sufficiently
large $\Lambda$, since it diverges as $\ln (\Lambda^2/r)$. However,
the subtraction removes precisely this dominant growing term. Given
that the function $f''(r)$ is negative (besides finite), $f'_{\rm
ren}(r) < f'_{\rm ren}(r_0)$ if $r > r_0$.  This could induce one to
try to make the subtraction at the highest $r_0$ possible.  This might
be the idea behind the procedure proposed in Ref.~\cite{FoLa},
where it is demanded that $\lim_{r\rightarrow\infty}f_{\rm ren}(r) =
0$.  However, this prescription implies subtracting from $f$ a
function that is not a {\em first degree polynomial} in $r$, 
unlike in standard renormalization prescriptions, as exposed 
here (see also \cite{MS}).

Similar but more complicated expressions are obtained for 
the positive curvature case, the $D$-dimensional sphere $S^D$. 
In this case, one must also consider that for $r=0$ the zero mode 
must be removed from the discrete spectrum, as done for $D=2$ in 
Ref.~\cite{I2}. This subtraction, however, does not spoil positivity 
of the second term in Eq.~(\ref{pos}). 

Let us clarify the role of the trace anomaly, ${\cal A}$.  It is well
known that renormalization of the free action on a curved
even-dimensional spacetime demands the presence of a term proportional
to the curvature. It absorbs a logarithmic divergence that appears in
addition to the logarithmically divergent term proportional to $m^2$
present on the plane \cite{BirDav}. Thus, the logarithmic derivative
of $W$ with respect to the scale $R$ has two components: the
stress-tensor trace on the plane $\Theta$ plus an additional part,
independent of $m$ and proportional to $R^{-D}$, the {\em trace} or
{\em conformal anomaly}.  The alert reader may have noticed that the
original form of the $R$-monotonicity theorem (\ref{OsbR}) in
Ref.~\cite{OsbS} has $R^D\,\langle T_a^a\rangle$ in place of
$R^{D}\langle\Theta \rangle$, but it does not matter because the
difference is a constant. Nevertheless, adding this constant would
have been a convenient normalization for the critical value of the
monotonic quantity, had it been precisely $c(u) = R^D\,\langle
T_a^a\rangle$, as proposed in Refs.~\cite{Cardy,FoLa}: it would make
it proportional to the {\em conformal central charge}.  However, since
the correct monotonic quantity is rather $\tilde{c}(u) =
u^{-y}\,R^{D}\langle\Theta \rangle$, adding the conformal anomaly
would result in a divergence at the critical point.

Let us say a few words about the flat space limit $R
\rightarrow\infty$.  To take this limit, the function $f$ is no longer
appropriate, and one must instead consider a local quantity, such as
$R^{y-D}\,f(mR)= -V_{D-1}\,m^{-y}\,\langle\Theta\rangle$. Thus, for
the massive free field theory in $D=3$, $\lim_{R
\rightarrow\infty}R^{y-D}\,f(mR)= m$. In contrast, for $D=2$, $R^{0}
f(mR) = f(mR)$ diverges logarithmically as $R \rightarrow\infty$, as
deduced from the corresponding asymptotic expansion \cite{I2}.  It is
because $R$ plays the role of an IR cutoff, and $f(0)$ on the plane is
IR divergent, as well as UV divergent. The solution is to subtract at
$r_0 \neq 0$ before taking the limit, which will depend on $m_0$ and,
therefore, one cannot construct a universal quantity.  The same
problem exists in $D=4$, even though in this case one should not give
particular value to the point $m_0 = 0$, as remarked above.  Let us
note, in passing, that the leading terms of the asymptotic expansion
of $f(u)$ yield the flat space limit and, furthermore, for even
dimension, the sub-leading term yields the conformal anomaly \cite{I2}.

In conclusion, the monotonicity theorems for the relative entropy or 
for the stress-tensor trace are contained in Eqs.~(\ref{pos}). In field 
theory, $\langle (f_\l-\langle f_\l\rangle)^2\rangle$ is proportional 
to the integral of the stress-tensor trace correlation, which only 
converges if $y$, the dimension of the coupling constant $\l$, satisfies 
$y > D/2$. Therefore, only under this condition is the irreversibility 
theorem unambiguous. 
However, even in this case, the function $f$ associated
to the stress-tensor trace is ambiguous (non-universal), being defined only 
up to a constant, whereas 
the relative entropy is unambiguous (universal). After renormalization, 
the ambiguity of $f$ is realized as a dependence on $m_0$, which is
the renormalization point in the simple scheme used here. 
Setting $m_0 = 0$ achieves a kind
of universality, in the sense that no additional scale remains, but
it may not be realizable, as occurs for free field theory on the
plane. 
The case $y > D/2$ covers many of the critical
models of Statistical Mechanics, e.g., the $3D$ Ising model
universality class, with $y = 1.59$ \cite{Z-J}.  
When $y \leq D/2$ (in particular, for bosonic free-field theory in $D=4$)
the integral in the right-hand side of Eq.~(\ref{pos2}) is UV divergent 
and must be renormalized, in general spoiling its positivity, so the
irreversibility theorem is itself ambiguous and may only hold in a particular
renormalization scheme. 
Accordingly, $f$ needs to be subtracted twice.  
Of course, the problem of the divergence of that integral
also affects the relative entropy, 
which becomes non universal, requiring one additional
subtraction further to those implied in its definition (\ref{Legtrans}).
Hence, it is doubtful whether one can assign an unambiguous meaning 
to RG irreversibility for $y \leq D/2$. 

Since irreversibility in terms of the stress-tensor trace 
or in terms of the relative
entropy are essentially equivalent, one may wonder which formulation
is better.  From a physical point of view, the theorem for the
relative entropy has more content, being related to important notions
in Information Theory \cite{I-OC,Pres}, while from a mathematical point of
view, $\langle\Theta\rangle$ is simpler to calculate and, in fact, to
calculate $S_{\rm rel}$ one must calculate it before (as in Eq.~(\ref{S})).
\vskip 2mm
I thank Hugh~Osborn for a conversation and for patiently explaining to me
tricky points on some calculations in Ref.~\cite{OsbS}

\end{multicols}

\begin{thebibliography}{99}
\bibitem{MS} In Ref.~\cite{OsbS}, Appendix D, 
$\langle\Theta \rangle = m^2\,\langle\phi^2 \rangle$ is computed for 
free theories in various dimensions 
with dimensional regularization and minimal subtraction.
\bibitem{xi} In Ref.~\cite{OsbS} the authors consider the scalar
conformally coupled field theory, with curvature coupling $\xi =
(D-2)/[4(D-1)]$, whereas here we set $\xi =0$. Consequently, for $D =
3,4$ some inessential numbers differ.
\end{thebibliography}
\end{document}